\documentstyle[aps,prl,epsfig,multicol]{revtex}
\begin{document}
\title{Electric Field Induced Kondo Tunneling Through Double Quantum Dot}

\author{M.N. Kiselev$^1$, K. Kikoin$^2$ and L.W.Molenkamp$^3$}
\address{
$^1$Institut f\"ur Theoretische Physik, $^3$Physikalisches Institut ($EP\;3$), Universit\"at W\"urzburg,
D-97074 W\"urzburg, Germany\\
$^2$Ben-Gurion University of the Negev, Beer-Sheva 84105, Israel}
\date{\today}


\twocolumn[\hsize\textwidth\columnwidth\hsize\csname@twocolumnfalse\endcsname

\maketitle
\begin{abstract}
It is shown that the resonance Kondo tunneling  through 
a double quantum dot (DQD) with {\it  even} occupation and
{\it singlet} ground state may arise at a strong bias, which compensates 
the energy 
of singlet/triplet excitation. Using the renormalization group technique 
we derive scaling equations and calculate the differential conductance as 
a function of an auxiliary dc-bias for parallel DQD.
\\
\mbox{}\\
PACS  numbers: 72.10.-d, 72.10.Fk, 72.15.Qm, 05.10.Cc \\
\end{abstract}
]

Many  fascinating  collective effects, which exist in 
strongly correlated electron systems (metallic
compounds containing transition and rare-earth elements) may be 
observed also in artificial nanosize devices
(quantum wells, quantum dots, etc). Moreover,  fabricated 
nanoobjects provide unique possibility to create
such conditions for observation of many-particle phenomena, 
which by no means may be reached in "natural" conditions. 
Kondo effect (KE) is one of such phenomena. 
It was found theoretically \cite{Glazr88b} 
and observed experimentally \cite{Gogo99} that the charge-spin 
separation in low-energy excitation 
spectrum of quantum dots under strong Coulomb blockade manifests 
itself as a resonance Kondo-type tunneling through a dot with 
odd electron occupation ${\cal N}$ (one unpaired spin 1/2).
This resonance tunneling through a quantum dot connecting two 
metallic reservoirs (leads) is an analog of resonance spin scattering 
in metals with magnetic impurities. 
A Kondo-type tunneling may be observed under conditions 
which do not exist in conventional metallic 
compounds. In particular, Kondo effect  survives in essentially 
non-equilibrium state when the
 strong bias $eV \gg T_K$ is applied between the leads 
\cite{Meir} ($T_K$ is the Kondo temperature which
determines the energy scale of low-energy spin excitations 
in a quantum dot). The KE may be observed as a 
dynamical phenomenon in strong time dependent electric field \cite{Goldin}, 
it may arise at finite frequency under light illumination \cite{Kex}. 
Even the net zero spin of isolated quantum
dot (even ${\cal N}$) is not the obstacle 
for the resonance Kondo tunneling. In this case it may be
observed in specific types of double quantum dots (DQD) \cite{KA01} 
or induced by strong magnetic field
\cite{Magn} whereas in conventional metals magnetic field only 
suppresses the Kondo scattering. The latter
effect was also observed experimentally \cite{Sasa}.

As was noticed in \cite{KA01,KA02}, quantum dots with even ${\cal N}$ possess
the dynamical symmetry $SO(4)$ of spin rotator in the Kondo tunneling regime,
provided the low-energy part of its spectrum is formed by a singlet-triplet
(ST) pair, and all other excitations are separated from the ST manifold by
a gap noticeably exceeding the tunneling rate $\gamma$. A DQD with 
even ${\cal N}$ in a side-bound configuration
where two wells are coupled by the tunneling $v$ and only 
one of them (say, $l$) is coupled to metallic 
leads $(L,R)$ is a simplest system satisfying this condition
\cite{KA01}. Such system was realized experimentally in Ref.\cite{mol95}.

In the present paper one more unusual property of DQDs with even ${\cal N}$
is revealed. 
It is shown that in the case when the ground state is singlet $|S\rangle$
and the ST gap $\delta \gg T_K$, a Kondo resonance 
channel arises under a strong bias $eV$ comparable with $\delta.$
The channel opens at 
$|eV - \delta|< T_K$, and the Kondo temperature 
is determined by the 
{\it non-diagonal} component 
$J_{ST}=\langle T|J|S\rangle$ of effective exchange 
induced by the electron tunneling through DQD (Fig. 1b). 
\begin{figure}
\begin{center}
\epsfxsize42mm
\epsfysize28mm
\epsfbox{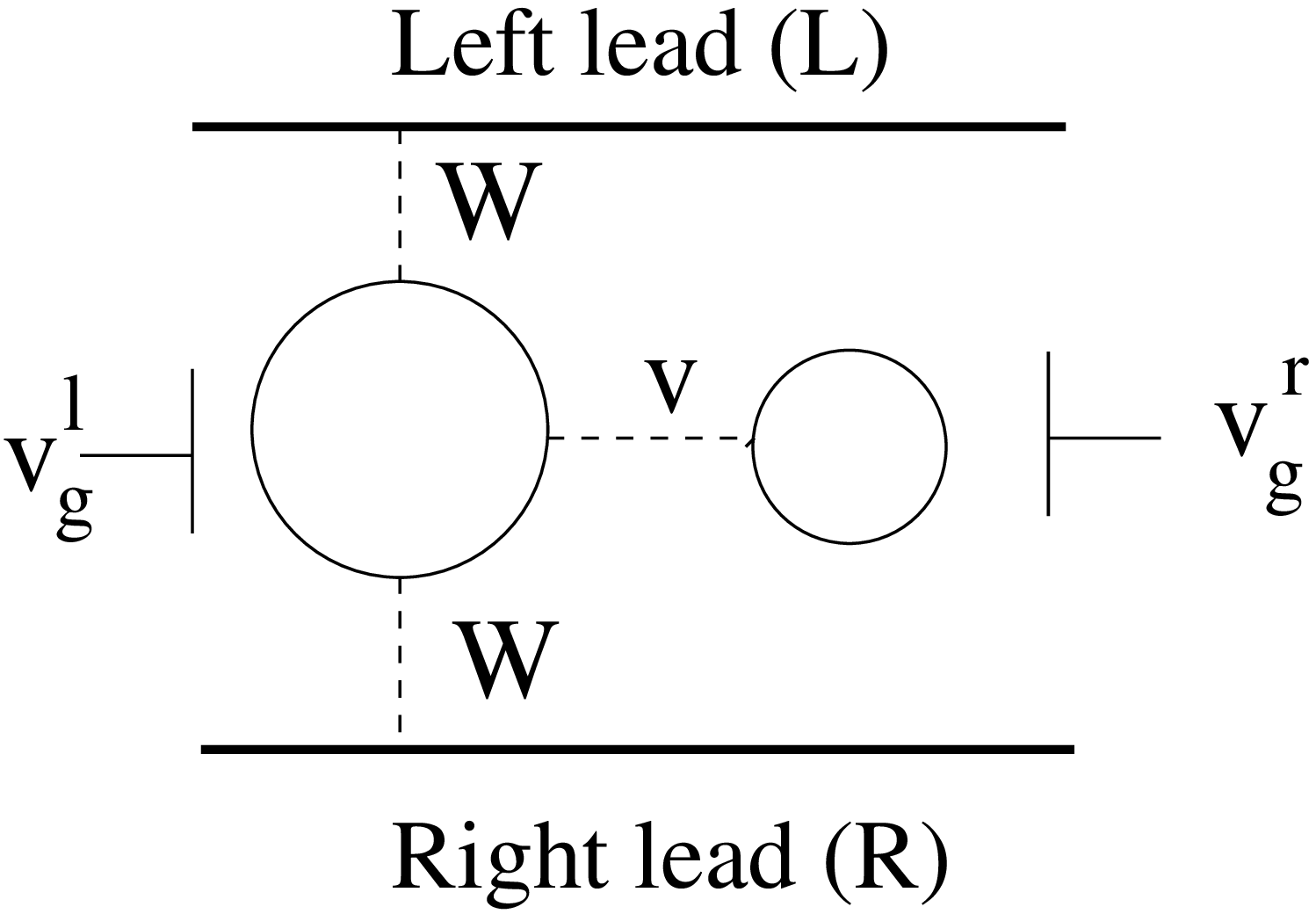}
\epsfxsize42mm
\epsfbox{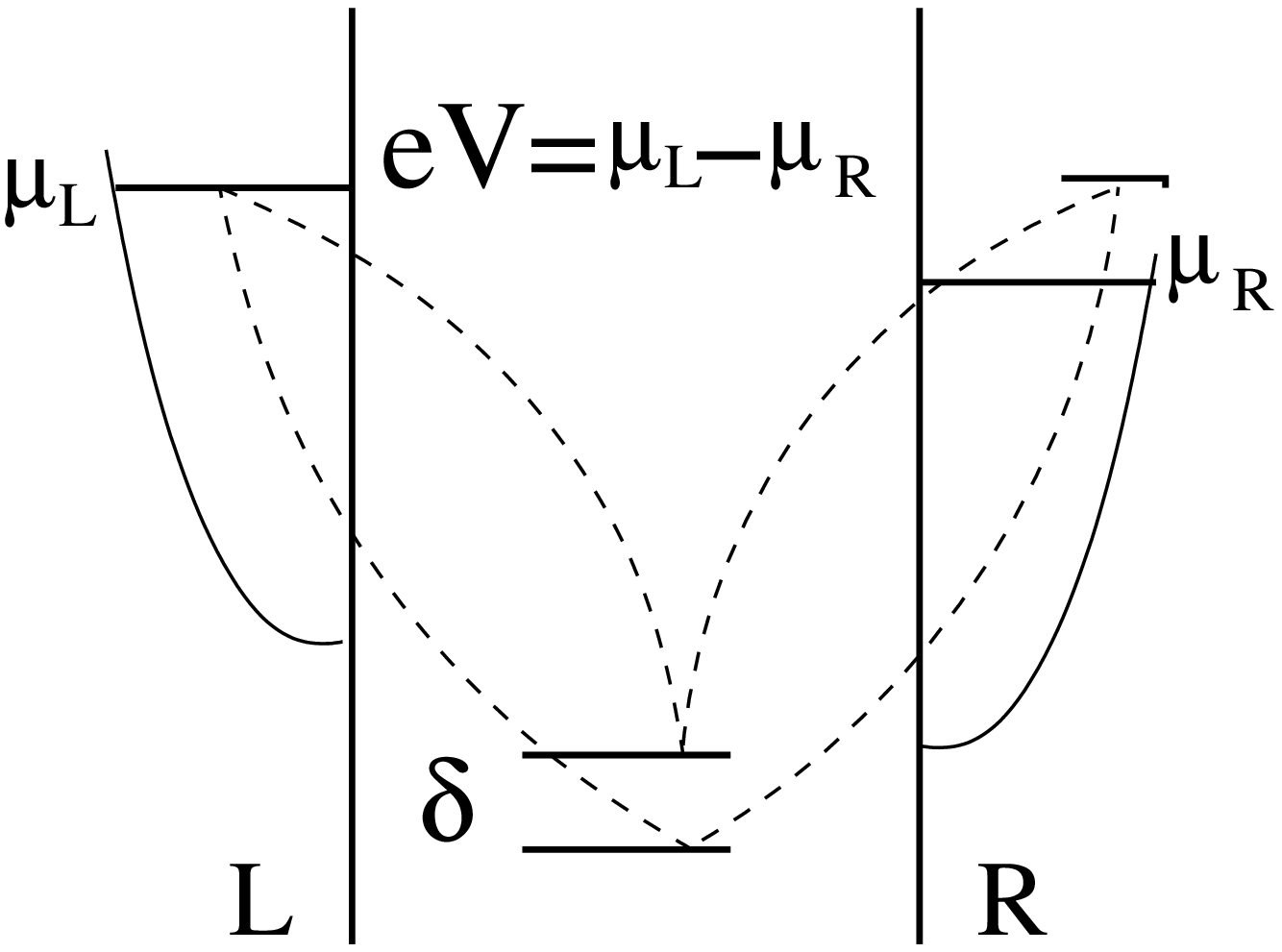}
a)\hspace*{40mm} b)\\
\caption{(a) Double quantum dot in a side-bound configuration
(b) co-tunneling processes 
in biased DQD responsible for the resonance Kondo tunneling.}
\end{center}
\end{figure}
\vspace*{-5mm}
The basic properties of symmetric
DQD occupied by even number of electrons ${\cal N}=2n$ under strong 
Coulomb blockade in each well are 
manifested already in the simplest case $n=1$, which is considered below. 
Such DQD is an artificial 
analog of a hydrogen molecule ${\rm H_2}$. If the inter-well Coulomb blockade 
$Q$ is strong enough, one has ${\cal N}=n_l+n_r,$ $n_l=n_r=1,$  
the lowest states of DQD are singlet and triplet and
the next levels are separated from ST pair by a charge transfer gap $\sim Q$.
We assume that both wells are neutral at $n_{l,r}=1$. 
Then the effective inter-well exchange responsible for the singlet-triplet
splitting arises because of 
tunneling $V$ between two wells, $J=v^2/Q=\delta$. It is convenient
to write the effective spin Hamiltonian of isolated DQD in the form
\begin{equation}
H_{d}=E_S|S\rangle\langle S|+\sum_{\eta}E_T|T\eta\rangle\langle T\eta| \equiv 
\sum_{\Lambda=S,T\eta}E_{\Lambda}X^{\Lambda\Lambda}
\label{1.1}
\end{equation}
where $X^{\Lambda\Lambda'}=|\Lambda\rangle\langle \Lambda'|$ is a Hubbard
configuration change operator (see, e,g, \cite{Hewson}),
$E_T=E_S+\delta$, $\eta=\pm,0$ are three projections of $S=1$ vector. 
Two other terms completing the Anderson Hamiltonian, which describes 
the system shown in Fig.1a, are
$$
H_{b}+H_{t}=
$$
\begin{equation}
\sum_{k\alpha\sigma}\epsilon_{k\alpha}
c^{\dagger}_{k\alpha\sigma}c_{k\alpha\sigma} + 
W\sum_{\Lambda \lambda }\sum_{k\sigma}\left(
c_{kl\sigma}^{\dagger }X^{\lambda \Lambda }+ H.c.
\right).   
\label{2.1}
\end{equation}
The first term describes metallic electrons in the leads and the second one stands for tunneling 
between the leads and the DQD. Here $\alpha=L,R$ marks 
electrons in the left and right lead, respectively, 
the bias $eV$ is applied to the left lead, so that the chemical potentials are 
$\mu_{FL}=\mu_{FR}+eV$,~ $W$ is the tunneling amplitude for the well $l$,
$|\lambda\rangle$ are one-electron states of DQD, 
 which arises after escape of an electron with spin projection
$\sigma$ from DQD in a state $|\Lambda\rangle$.

In case of strong Coulomb blockade $Q\gg \gamma$, the non-equilibrium
repopulation of DQD is a weak effect and one may 
start with a second order perturbation 
calculation in tunneling amplitude known as the Schrieffer-Wolff (SW) 
transformation \cite{Hewson}, 
where both leads are considered as independent
subsystems.  
As is shown in Refs. \cite{KA01,KA02} 
the SW transformation being applied to a
spin rotator results in the following 
effective spin Hamiltonian 
\begin{equation}
H_{int}=\sum_{\alpha\alpha^\prime}[
(J^{TT}_{\alpha\alpha^\prime}{\bf S} +  J^{ST}_{\alpha\alpha^\prime}
{\bf P})\cdot {\bf s}_{\alpha\alpha^\prime})
+J^{SS}_{\alpha\alpha^\prime}X^{SS}n_{\alpha\alpha^\prime}] 
\label{2.2}
\end{equation}
Here ${\bf s}_{\alpha\alpha^\prime}$$=$$
\sum_{kk'}c^{\dagger}_{k\alpha\sigma}\hat{\tau}c_{k'\alpha'\sigma'}$,~
$n_{\alpha\alpha^\prime}$$=$$
\sum_{kk'}c^{\dagger}_{k\alpha\sigma}\hat{1}c_{k'\alpha'\sigma}$,
$\hat{\tau}$,~ $\hat{1}$ are the Pauli matrices and unity matrix respectively.
The effective exchange constants are
$$
J^{\Lambda\Lambda'}_{\alpha\alpha^\prime} \approx \frac{W^2}{2}\left(
\frac{1}{\epsilon_{F\alpha}-(E_S/2+\delta)} +
\frac{1}{\epsilon_{F\alpha^\prime}-(E_S/2+\delta)}
\right)
$$
Two vectors ${\bf S}$ and ${\bf P}$  with spherical components
\begin{eqnarray}
\nonumber
S^+  = \sqrt{2}\left(X^{10}+X^{0-1}\right),~ 
S^-  =  \sqrt{2}\left(X^{01}+X^{-10}\right),\\
\nonumber
S_z  =  X^{11}-X^{-1-1},\,\;\;\;
P_z  =  -\left(X^{0S}+X^{S0}\right),\\
P^+  = \sqrt{2}\left(X^{1S}-X^{S-1}\right),
P^-  = \sqrt{2}\left(X^{S1}-X^{-1S}\right).
\label{SP}
\end{eqnarray}
obey the commutation relations of $o_4$ algebra
\[
[S_j,S_k]  = ie_{jkl}S_l,~[P_j,P_k]=ie_{jkl}S_l,~ [P_j,S_k]=ie_{jkl}P_l
\]
($j,k,l$ are Cartesian coordinates, $e_{jkl}$ is a Levi-Chivita tensor). 
These vectors are orthogonal, ${\bf S\cdot P} = 0,$ and the Casimir operator
is ${\bf S}^2+ {\bf P}^2 =3.$ Thus, the singlet state is involved
in spin scattering via the components of the vector ${\bf P}$. 

We use $SU(2)$-like semi-fermionic representation for $S$ operators 
\cite{popov,kis}
$$
S^+  =   \sqrt{2}(f_0^\dagger f_{-1}+f^\dagger_{1}f_0),\;\;\;
S^-  =  \sqrt{2}(f^\dagger_{-1}f_0+ f_0^\dagger f_{1}),
$$
\begin{equation}
S_z  =  f^\dagger_{1}f_{1}-f^\dagger_{-1}f_{-1},
\label{spin}
\end{equation}
where $f^\dagger_{\pm}$ are creation operators for fermions with spin
``up'' and ``down'' respectively,
whereas $f_0$ stands for spinless fermion \cite{popov,kis}. 
This representation can be 
generalized for $SO(4)$ group by introducing another spinless fermion $f_s$
to take into consideration the  singlet state. As a result, the $P$-operators 
are given by the following equations:
$$
P^+ =  \sqrt{2}(f^\dagger_{1} f_s -  f_s^\dagger f_{-1}),\;\;\;
P^- =  \sqrt{2}(f_s^\dagger f_{1} - f^\dagger_{-1}f_s),
$$
\begin{equation}
P^z  =  -( f_0^\dagger f_s + f_s^\dagger f_0). 
\label{proj}
\end{equation}
The Casimir operator ${\bf S}^2+{\bf P}^2=3$ transforms to the local constraint
$\sum_{\Lambda=\pm,0,s}f^\dagger_\Lambda f_\Lambda=1$.

The spin Hamiltonian is now given by
\begin{equation}
H_{int}=\sum_{kk',\alpha\alpha'=L,R}J^S_{\alpha\alpha'}
f_s^\dagger f_s c^\dagger_{k\alpha\sigma}c_{k'\alpha'\sigma}
\label{hint}
\end{equation}
$$
+\sum_{kk',\alpha\alpha'\Lambda\Lambda'}\left(J^T_{\alpha\alpha'}
\hat S^d_{\Lambda \Lambda'}
+J^{ST}_{\alpha\alpha'}\hat P^d_{\Lambda \Lambda'}\right)\tau^d_{\sigma\sigma'}
c^\dagger_{k\alpha\sigma}c_{k'\alpha'\sigma'}f_\Lambda^\dagger f_{\Lambda'}
$$
where $\hat S^d$ and $\hat P^d$ ($d$$=$$x$,$y$,$z$) are $4\times 4$ 
matrices  defined by relations (\ref{SP}) - (\ref{proj})
and $J^S=J^{SS}$, $J^T=J^{TT}$ and $J^{ST}$ are singlet, triplet and singlet-triplet 
coupling SW constants, respectively.

To develop the perturbative approach for $T>T_K$ 
we introduce
the temperature Green's functions (GF) for electrons in a dot, 
${\cal G}_{\Lambda}(\tau)=-\langle T_\tau f_\Lambda(\tau) f^\dagger_\Lambda(0)\rangle,$
and GF of left (L) and right (R) electrons in the leads 
$G_{L,R}(k,\tau)=-\langle T_\tau c_{L,R \sigma}(k,\tau) 
c^\dagger_{L,R\sigma}(k,0)\rangle$.
Performing a Fourier transformation in imaginary time for bare GF's, 
we come to following expressions:
$$
G^0_{k\alpha}(\epsilon_n)=(i\epsilon_n -\epsilon_k+\mu_{L,R})^{-1},
$$
$$
{\cal G}^0_{\eta}(\omega_m)=(i\omega_m -E_T)^{-1},\;\;\; \eta=-1,0,1
$$
\begin{equation}
{\cal G}^0_{s}(\epsilon_n)=(i\epsilon_n -E_S)^{-1},
\label{GF}
\end{equation}
with $\epsilon_n=2\pi T(n+1/2)$ and $\omega_m=2\pi T(m+1/3)$ 
\cite{popov,kis}. 
The first leading and next to leading parquet diagrams are shown on Fig.2.

In equilibrium state $eV=0$ the elastic Kondo tunneling arises 
only provided $T_K \gg\delta$ in accordance with the theory of two-impurity
Kondo effect \cite{KA02,Varma}. Now we will show that 
in the opposite limit $T_K \ll\delta$
the elastic channel emerges at
$eV\approx \delta$. 

Corrections to the 
singlet vertex $\Gamma(\omega,0; \omega,eV)$ are calculated using 
an analytical continuation of GF's to the real axis 
$\omega$ and taking into 
account the shift of the chemical potential in the left lead. 
In a weak coupling regime $T>T_K$
the leading non-Born contributions to the tunnel current are determined
by the diagrams of Fig. 2 b-e.
\begin{figure}
\begin{center}
\epsfxsize80mm
\epsfysize50mm
\epsfbox{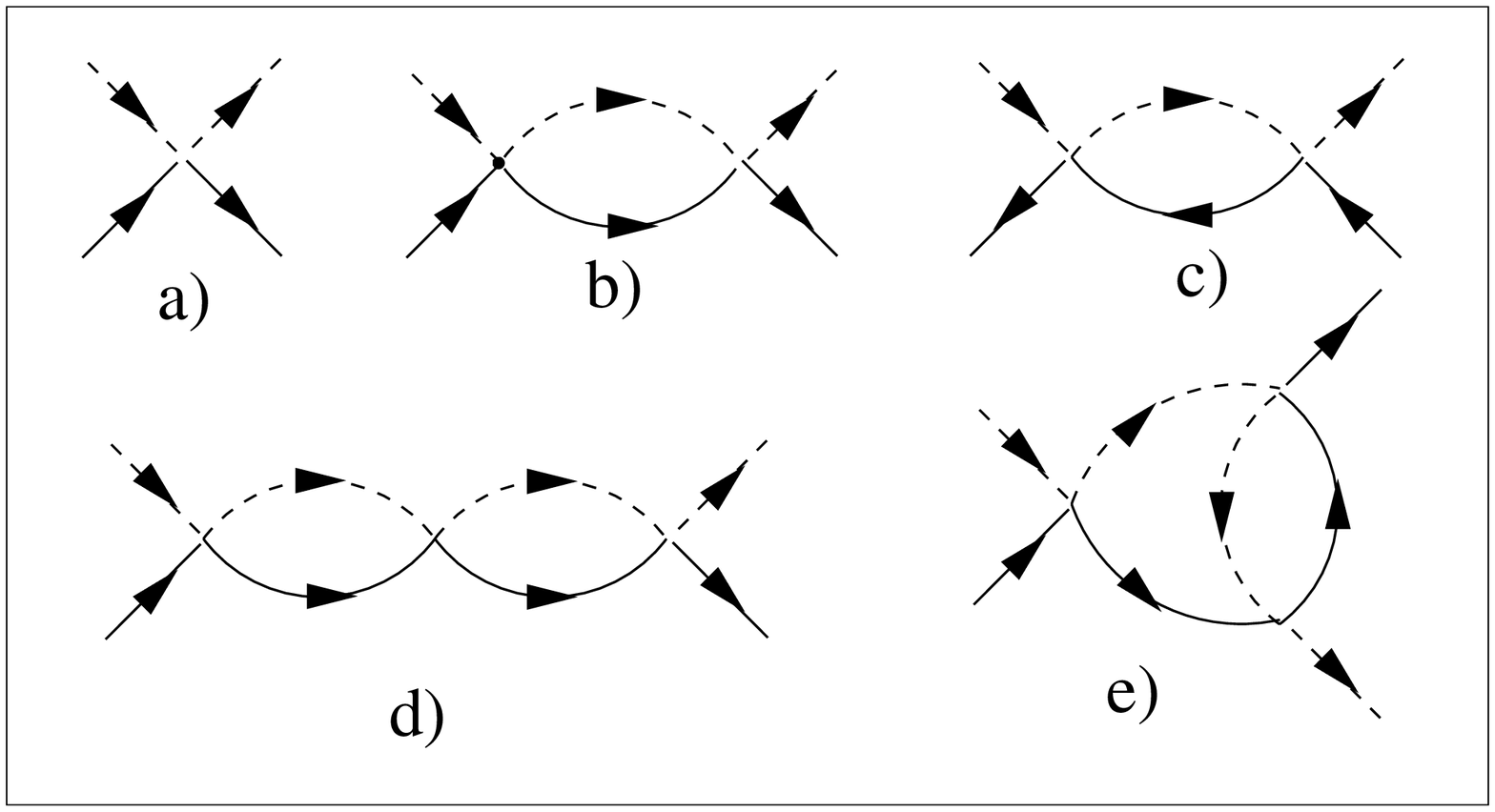}
\mbox{}\\
\mbox{}\\
\caption{Leading (b,d) and next to leading (c,e) parquet diagrams
determining renormalization of $J^S (a)$. Solid lines denote electrons in the leads. 
Dashed lines stand for electrons in the dot.}
\end{center}
\end{figure}
\vspace*{-5mm}
 The effective vertex shown in Fig. 2b
is given by the following equation
\begin{equation}
\Gamma_{LR}^{(2b)}(\omega)=J_{LL}^{ST}J_{LR}^{TS}\sum_{\bf k}
\frac{1-f(\epsilon_{kL}-eV)}
{\omega - \epsilon_{kL} + \mu_L - \delta}
\end{equation}
Changing the variable $\epsilon_{kL}$ for $\epsilon_{kL}-eV$ one finds
that $\Gamma_{LR}^{(2b)}(\omega)\sim 
J_{LL}^{ST}J_{LR}^{TS} \nu \ln~(D/{\rm max}\{\omega, (eV-\delta),T\})$. 
Here $D \sim \varepsilon_F$ is a cutoff energy determining
effective bandwidth, $\nu$ is a density of states on a Fermi level and
$f(\varepsilon)$ is the Fermi function. 
Therefore, under  condition $|eV-\delta|\ll \max[eV,\delta]$ 
this correction does not depend on $eV$ and becomes quasielastic.

Unlike the diagram Fig. 2b, 
its "parquet counterpart" term Fig. 2c contains  $eV+\delta$ 
in the argument of the Kondo logarithm:
\begin{equation}
\Gamma_{LR}^{(2c)}(\omega)=J_{LL}^{ST}J_{LR}^{TS}\sum_{\bf k}
\frac{f(\epsilon_{kL}-eV)}
{\omega - \epsilon_{kL} + \mu_L + \delta}
\end{equation}
At $eV\sim\delta \gg T,\omega$ this contribution is estimated 
as $\Gamma_{LR}^{(2c)}(\omega)\sim 
J_{LL}^{ST}J_{LR}^{TS}\nu \ln~(D/(eV+\delta))
\ll \Gamma_{LR}^{(2b)}(\omega)$.

Similar estimates for diagrams Fig.2d and 2e give
$$\Gamma_{LR}^{(2d)}(\omega)\sim
J_{LL}^{ST}J_{LL}^{T}J_{LR}^{TS}\nu^2\ln^2~(D/{\rm max}\{\omega, (eV-\delta),T\})
$$
$$\Gamma_{LR}^{(2e)}(\omega)\sim 
J_{LL}^{ST}J_{LL}^{T}J_{LR}^{TS}\nu^2\ln~(D/{\rm max}\{\omega, (eV-\delta),T\})\times
$$
\begin{equation}
\times \ln~(D/{\rm max}\{\omega, eV, T\}.
\label{2nd}
\end{equation}
Then $\Gamma_{LR}^{(2e)}(\omega)\ll\Gamma_{LR}^{(2d)}(\omega)$ 
at $eV \to \delta$.

Thus, the 
Kondo singularity is restored in strongly non-equilibrium conditions when the
energy loss $\delta$ in a singlet-triplet excitation is compensated by the
external voltage applied to the lead, but the leading sequence of 
most divergent diagrams degenerates in this case from a parquet to a 
ladder series.

Following the poor man's scaling approach, we derive the system of coupled 
renormalization group (RG) equations for (\ref{hint}). The equations for LL 
co-tunneling are:
\begin{figure}
\begin{center}
\epsfxsize80mm
\epsfysize50mm
\epsfbox{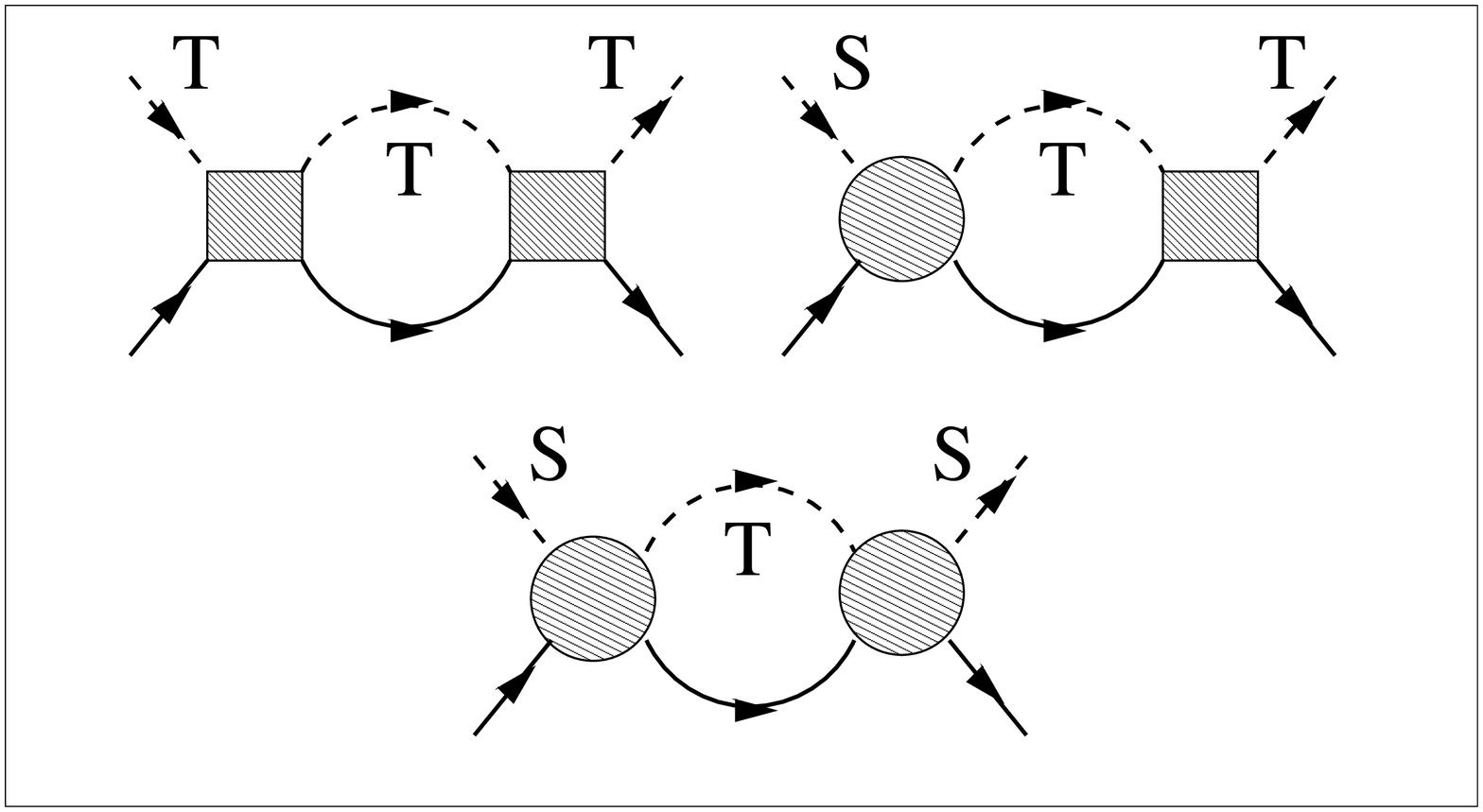}
\mbox{}\\
\mbox{}\\
\caption{Irreducible diagrams contributing to RG equations. Hatched boxes and circles stand for 
triplet-triplet and singlet-triplet vertices respectively. Notations for lines are the same as in Fig.2}
\end{center}
\end{figure}
\vspace*{-5mm}
\begin{equation}
\frac{d J^T_{LL}}{d \ln D}=-\nu (J_{LL}^T)^2,\;\;\;
\frac{d J^{ST}_{LL}}{d \ln D}=-\nu J_{LL}^{ST} J_{LL}^T,
\label{ll}
\end{equation}
The scaling equations for $J_{LR}^\Lambda$ are as follows:
$$
\frac{d J^T_{LR}}{d \ln D}=-\nu J_{LL}^T J_{LR}^T,\;\;\;
\frac{d J^{ST}_{LR}}{d \ln D}=-\nu J_{LL}^{ST} J_{LR}^T,
$$
\begin{equation}
\frac{d J^S_{LR}}{d \ln D}=\frac{1}{2}\nu\left(J_{LL,+}^{ST}J_{LR,-}^{TS}
+\frac{1}{2}J_{LL,z}^{ST}J_{LR,z}^{TS}\right).
\label{rg}
\end{equation}
One-loop diagrams corresponding to the poor man's scaling procedure are shown 
in Fig. 3. 
To derive these equations we collected only terms 
$\sim (J^T)^n\ln^{n+1}(D/T)$ neglecting 
contributions containing $\ln[D/(eV)]$. The analysis of RG equations beyond 
the one loop approximation 
will be published elsewhere.

The solution of the system (\ref{rg}) reads as follows:
$$
J^T_{\alpha,\alpha'}=\frac{J^T_{0}}{1-\nu J^T_{0}\ln(D/T)},\;\;\;
J^{ST}_{\alpha,\alpha'}=\frac{J^{ST}_{0}}{1-\nu J^T_{0}\ln(D/T)},
$$
\begin{equation}
J^S_{LR}=J^S_{0}-\frac{3}{4}\nu (J^{ST}_0)^2\frac{\ln(D/T)}{1-\nu J^T_{0}\ln(D/T)}.
\label{srg}
\end{equation}
Here $\alpha=L$, $\alpha'=L,R$.
The Kondo temperature is determined by triplet-triplet processes only. 
It is given by 
$T_K=D\exp[-1/(\nu J^T_0)]$.

The differential conductance $G(eV,T)/G_0$ is the universal function of 
two parameters $T/T_K$ and $V/T_K$ (see Fig. 4), $G_0=e^2/\pi \hbar$:
\begin{equation}
G/G_0 \sim \ln^{-2}\left(\max[(eV-\delta),T]/T_K\right)
\label{dcond}
\end{equation}
Finite decoherence rate $\hbar/\tau_d$ effects discussed in \cite{neq} in a context of 
strongly-nonequilibrium transport through QD with $S=1/2$ do not arise in our case. 
According to the 
Non-Crossing Approximation (NCA) description, the origin of $\tau_d$ is inelastic
spin relaxation of Kondo state. In the problem considered the ground state is $S=0$ singlet
and the spin-relaxation is absent. A Kondo-channel arises only in {\it
virtual} 
states of L-R co-tunneling.
Repopulation effects at $eV>\delta$ should result in  asymmetry of Kondo-peak
\cite{neq}, 
but
this effect is beyond our quasi equilibrium approach.
\begin{figure}
\begin{center}
\epsfxsize75mm
\epsfbox{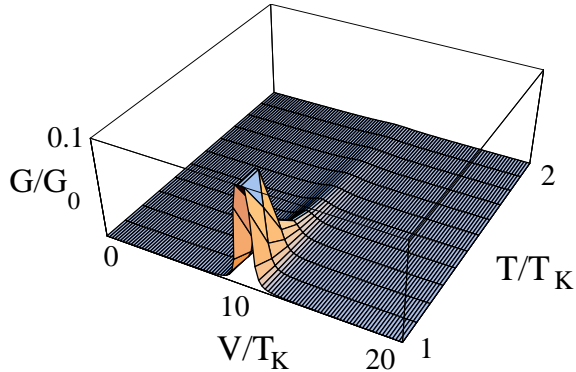}
\mbox{}\\
\mbox{}\\
\caption{The Kondo conductance as a function of dc-bias $eV/T_K$ and $T/T_K$.
The singlet-triplet splitting $\delta/T_K=10$.}
\end{center}
\end{figure}
\vspace*{-5mm}
Thus, we have shown that the tunneling through singlet DQDs with 
$\delta\gg T_K$ 
exhibits a peak in 
differential conductance at $eV\approx \delta$ 
instead of the usual zero bias Kondo anomaly (see Fig. 4) which arises in 
the opposite limit, $\delta < T_K$.
Therefore, in this case the Kondo effect in DQD is induced by 
a strong external bias. 
The scaling equations (\ref{rg}), (\ref{srg}) can also be derived in 
Schwinger-Keldysh formalism (see \cite{kis} and also \cite{neq}) by
applying the ``poor man's scaling'' approach directly to the dot conductance
\cite{Goldin}. The detailed analysis of the model (\ref{hint}) 
in a real-time formalism will be presented elsewhere.

We discuss yet another possible experimental realization 
of resonance Kondo tunneling driven 
by external electric field. Applying the alternate field $V=V_{ac}\cos(\omega t)$ to the 
parallel DQD, one takes into consideration two effects, namely (i) enhancement of 
Kondo conductance by tuning the amplitude of ac-voltage to satisfy the condition 
$|eV_{ac}-\delta| \ll T_K$ and (ii) spin decoherence effects due to finite decoherence rate  \cite{Goldin}. 
One can expect that if the decoherence rate $\hbar/\tau \gg T_K,$
\begin{equation}
G_{peak}/G_0 \sim \ln^{-2}\left(\hbar/\tau T_K\right)
\end{equation}
whereas in the opposite limit $\hbar/\tau \ll T_K$,

\begin{equation}
G_{peak}=\overline{G(V_{ac}\cos[\omega t])}
\end{equation}
is averaged over a period of variation of ac bias. In this case the estimate 
(\ref{dcond}) is also valid. 

In conclusion, we have provided the first example of Kondo effect, which
exists {\it only} in non-equilibrium conditions. It is driven by external 
electric field in tunneling through a
quantum dot with even number of electrons, when the low-lying states are 
those of spin rotator. This is not too exotic situation because as a rule, 
a singlet ground state implies a triplet excitation. If the
ST pair is separated by a gap from other excitons, then  
tuning the dc-bias in such a way that applied voltage compensates  
the energy of triplet excitation,
one reaches the regime of Kondo peak in conductance. 
This theoretically predicted effect 
can be observed in dc- and ac-biased double quantum dots in parallel geometry.

This work is partially supported (MK) by the European Commission under LF project: Access 
to the Weizmann
Institute Submicron Center (contract number: HPRI-CT-1999-00069). The authors
are grateful to Y. Avishai, 
A. Finkel'stein, Y. Gefen, H. Kroha, A. Rosch and M.Heiblum  
for numerous useful discussions. The financial support of the Deutsche 
Forschungsgemeinschaft (SFB 410) is  gratefully 
acknowledged. The work of KK is supported by ISF grant.


\begin{thebibliography}{99}
\bibitem{add} Possibility  of additional Kondo peaks at $eV\gg T_K$ in strong magnetic field
was noticed also in the first and last papers of Ref.[7].
\end{thebibliography}
\end{document}